# Temperate Earth-sized planets transiting a nearby ultracool dwarf star

Michaël Gillon[1], Emmanuël Jehin[1], Susan M. Lederer[2], Laetitia Delrez[1], Julien de Wit[3], Artem Burdanov[1], Valérie Van Grootel[1], Adam J. Burgasser[4], Amaury H. M. J. Triaud[5], Cyrielle Opitom[1], Brice-Olivier Demory[6], Devendra K. Sahu[7], Daniella Bardalez Gagliuffi[4], Pierre Magain[1] & Didier Queloz[6]

[1]Institut d'Astrophysique et de Géophysique, Université de Liège, Allée du 6 Août 19C, 4000 Liège, Belgium.

[2]NASA Johnson Space Center, 2101 NASA Parkway, Houston, Texas, 77058, USA.

[3]Department of Earth, Atmospheric and Planetary Sciences, Massachusetts Institute of Technology, 77 Massachusetts Avenue, Cambridge, Massachusetts 02139, USA.

[4]Center for Astrophysics and Space Science, University of California San Diego, La Jolla, California 92093, USA.

[5]Institute of Astronomy, Madingley Road, Cambridge CB3 0HA, UK.

[6]Astrophysics Group, Cavendish Laboratory, 19 J J Thomson Avenue, Cambridge, CB3 0HE, UK.

[7]Indian Institute of Astrophysics, Koramangala, Bangalore 560 034, India.

**Star-like objects with effective temperatures of less than 2,700 kelvin are referred to as 'ultracool dwarfs'[1]. This heterogeneous group includes stars of extremely low mass as well as brown dwarfs (substellar objects not massive enough to sustain hydrogen fusion), and represents about 15 per cent of the population of astronomical objects near the Sun[2]. Core-accretion theory predicts that, given the small masses of these ultracool dwarfs, and the small sizes of their protoplanetary disk[3,4], there should be a large but hitherto undetected population of terrestrial planets orbiting them[5]—ranging from metal-rich Mercury-sized planets[6] to more hospitable volatile-rich Earth-sized planets[7]. Here we report observations of three short-period Earth-sized planets transiting an**

**ultracool dwarf star only 12 parsecs away. The inner two planets receive four times and two times the irradiation of Earth, respectively, placing them close to the inner edge of the habitable zone of the star[8]. Our data suggest that 11 orbits remain possible for the third planet, the most likely resulting in irradiation significantly less than that received by Earth. The infrared brightness of the host star, combined with its Jupiter-like size, offers the possibility of thoroughly characterizing the components of this nearby planetary system.**

TRAPPIST[9,10] (the TRansiting Planets and PlanestIsimals Small Telescope) monitored the brightness of the star TRAPPIST-1 (2MASS J23062928-0502285) in the very-near infrared (roughly 0.9 μm) at high cadence (approximately 1.2 minutes) for 245 hours over 62 nights from 17 September to 28 December 2015. The resulting light curves show 11 clear transit-like signatures with amplitudes close to 1% (Extended Data Figs 1, 2). Photometric follow-up observations were carried out in the visible range with the Himalayan Chandra 2-metre Telescope (HCT) in India, and in the infrared range with the 8-metre Very Large Telescope (VLT) in Chile and the 3.8-metre UK Infrared Telescope (UKIRT) in Hawaii. These extensive data show that nine of the detected signatures can be attributed to two planets, TRAPPIST-1b and TRAPPIST-1c, transiting the star every 1.51 days and 2.42 days, respectively (Fig. 1a ,b). We attribute the two additional transit signals to a third transiting planet, TRAPPIST-1d, for which 11 orbital periods—from 4.5 days to 72.8 days—are possible on the basis of non-continuous observations (Table 1). We cannot discard the possibility that the two transits attributed to planet TRAPPIST-1d originate instead from two different planets, but the consistency of their main parameters (duration, depth and impact parameter) as derived from their individual analyses does not favour this scenario.

TRAPPIST-1 is a well characterized, isolated M8.0 ± 0.5-type dwarf star[11] at a distance of 12.0 ± 0.4 parsecs from Earth as measured by its trigonometric parallax[12], with an age constrained to be more than 500 million years (Myr), and with a luminosity, mass and radius of 0.05%, 8% and 11.5% those of the Sun[13], respectively. We determined its metallicity to be solar through the analysis of newly acquired infrared spectra. The small size of the host star—only slightly larger than Jupiter—translates into Earth-like radii for the three discovered planets, as deduced from their transit depths. Table 1 presents the physical properties of the system, as derived through a global Bayesian analysis of the transit photometry (Fig. 1), including the *a priori* knowledge of its stellar properties, with an adaptive Markov-chain Monte Carlo (MCMC) code[14].

We can discard a non-planetary origin of the transit-like signals owing to several factors. The first factor is the high proper motion of the star (greater than 1″ per year), which allowed confirmation (through archival images) that no background source of significant brightness was located behind it in 2015. The second factor is that the star has no physical companion of stellar-like nature (star or brown dwarf), as demonstrated by high-resolution images, radial velocities and near-infrared spectroscopy. Together, these factors show that the signals do not originate from eclipses of larger bodies in front of a background or a physically associated stellar-like object blended with the ultracool target star. These factors also establish that the light from the target is not diluted by an unresolved additional stellar-like object, confirming that the measured transit depths reveal planetary radii of terrestrial sizes. Other factors include the significant age of the star[13], its moderate activity[15] and rotation period ($P_{rot}$ = 1.40 ± 0.05 days, as measured from our photometry), and its low level of photometric variability[16] (confirmed by our data), all of which are inconsistent with exotic scenarios based on ultrafast rotation of photospheric structures, or on occultations by circumstellar material of non-planetary origin (for example, disk patches or comets)[17].

Further confirmation of the planetary origin of the transits comes from, first, the periodicity of the transits of TRAPPIST-1b and TRAPPIST-1c, and the achromaticity of the transits of TRAPPIST-1b as observed from 0.85 μm (HCT) to 2.09 μm (VLT) (Fig. 1a); and second, the agreement between the stellar density measured from the transit light curves, $49.3^{+4.1}_{-8.3}\ \rho_{\odot}$, with the density inferred from the stellar properties, $55.3 \pm 12.1\ \rho_{\odot}$ (where $\rho_{\odot}$ is the density of the Sun).

The masses of the planets, and thus their compositions, remain unconstrained by these observations. The results of planetary thermal evolution models—and the intense extreme-ultraviolet (1−1,000 Å) emission of low-mass stars[18] during their early lives—make it unlikely that such small planets would have thick envelopes of hydrogen and/or helium gases[19]. Statistical analyses of sub-Neptune-sized planets detected by the Kepler spacecraft indicate that most Earth-sized planets in close orbit around solar-type stars are rocky[20,21]. Nonetheless, the paucity of material in the inner region of the protoplanetary disk of an ultracool dwarf would seem to challenge the *in situ* formation of rocky planets the size of Earth[6], favouring instead compositions dominated by ice-rich material originating from beyond the ice line[7]. Confirming this hypothesis will require precise mass measurements so as to break the degeneracy between the relative amounts of iron, silicates and ice[22]. This should be made possible by next-generation, high-precision infrared velocimeters able to measure the low-amplitude Doppler signatures (of one-half to a few metres per second) of the planets. Alternatively, the planets' masses could be constrained by measuring the transit timing variations (TTVs) caused by their mutual gravitational interactions[23], or by transit transmission spectroscopy[24].

Given their short orbital distances, it is likely that the planets are tidally locked—that is, that their rotations have been synchronized with their orbits by tidal interactions with the host star[25]. Planets TRAPPIST-1b and TRAPPIST-1c are not in the host star's habitable zone[10] (within 0.024 to 0.049 astronomical units (AU) of the star, as defined by one-dimensional models that are not adequate for modelling the highly asymmetric climate of tidally locked planets[26]). However, they have low enough equilibrium temperatures that they might have habitable regions—in particular, at the western terminators of their day sides[27] (Fig. 2 and Table 1). The main concern regarding localized habitability on tidally locked planets relates to the trapping of atmosphere and/or water on their night sides[28]. Nevertheless, the relatively large equilibrium temperatures of TRAPPIST-1b and TRAPPIST-1c would probably prevent such trapping[27]. In contrast, TRAPPIST-1d orbits within or beyond the habitable zone of the star, its most likely periods corresponding to semi-major axes of between 0.033 and 0.093 AU. We estimate tidal circularization timescales for TRAPPIST-1d (unlike for the two inner planets) to be more than 1 billion years (see the section "Dynamics of the system" in Methods). Tidal heating due to a non-zero orbital eccentricity could thus have a significant influence on the global energy budget and potential habitability of this planet[28].

The planets' atmospheric properties, and thus their habitability, will depend on several unknown factors. These include the planets' compositions; their formation and dynamical history (their migration and tides); the past evolution and present level of the extreme-ultraviolet stellar flux[29] (probably strong enough in the past, and perhaps even now, to significantly alter the planets' atmospheric compositions[30]); and the past and present amplitudes of atmospheric replenishment mechanisms (impacts and volcanism). Fortunately, the TRAPPIST-1 planets are particularly well suited for detailed atmospheric characterization—notably by transmission spectroscopy (Fig. 3)—because transit signals are inversely proportional to the square of the host-star radius, the latter being only ~12% of that of the Sun for TRAPPIST-1. Data obtained by the Hubble Space Telescope should provide initial constraints on the extent and composition of the planets' atmospheres. The next generation of observatories will then allow far more in-depth exploration of the atmospheric properties. In particular, data from the James Webb Space Telescope should yield strong constraints on atmospheric temperatures and on the abundances of molecules with large absorption bands, including several biomarkers such as water, carbon dioxide, methane and ozone.

**Acknowledgements** TRAPPIST is funded by the Belgian Fund for Scientific Research (FRS–FNRS) under grant FRFC 2.5.594.09.F, with the participation of the Swiss Fund for Scientific Research. The research leading to our results was funded in part by the European Research Council (ERC) under the FP/2007-2013 ERC grant 336480, and through an Action de Recherche Concertée (ARC) grant financed by the Wallonia-Brussels Federation. Our work was also supported in part by NASA under contract NNX15AI75G. The VLT/HAWK-I data used in this work were obtained in the Director Discretionary Time (DDT) program 290.C-5010. UKIRT is supported by NASA and operated under an agreement among the University of Hawaii, the University of Arizona, and Lockheed Martin Advanced Technology Center; operations are enabled through the cooperation of the East Asian Observatory. The facilities at the Indian Astronomical Observatory (IAO) and the Consortium for Research Excellence, Support and Training (CREST) are operated by the Indian Institute of Astrophysics, Bangalore. M.G., E.J. and V.V.G. are FRS–FNRS research associates. L.D. and C.O. are FRS–FNRS PhD students. We thank V. Mégevand, the ASTELCO telescope team, S. Sohy, V. Chantry, and A. Fumel for their contributions to the TRAPPIST project; the Infrared Telescope Facility (IRTF) operators B. Cabreira and D. Griep for assistance with the SpeX observations; UKIRT staff scientists W. Varricatt & T. Kerr, telescope operators S. Benigni, E. Moore and T. Carroll, and Cambridge Astronomy Survey Unit (CASU) scientists G. Madsen and M. Irwin for assistance with UKIRT observations; the European Southern Observatory (ESO) astronomers A. Smette and G. Hau for providing us with the best possible VLT data; and the staff of IAO (in Hanle) and CREST (in Hosakote) for making observations with the HCT possible. Ad.B. and D.B.G. are visiting astronomers at the IRTF, which is operated by the University of Hawaii under Cooperative Agreement NNX-08AE38A with NASA's Science Mission Directorate, Planetary Astronomy Program.

**Author Contributions** The TRAPPIST team (M.G., E.J., L.D., Ar.B., C.O. and P.M.) discovered the planets. M.G. leads the exoplanet program of TRAPPIST, set up and organized the ultracool-dwarf transit survey, planned and analysed part of the observations, led their scientific exploitation, and wrote most of the manuscript. E.J. manages the maintenance and operations of the TRAPPIST telescope. S.M.L. obtained the director's discretionary time on UKIRT, and managed, with E.J., the preparation of the UKIRT observations. L.D. and C.O. scheduled and carried out some of the TRAPPIST observations. L.D. and Ar.B. analysed some photometric observations. J.d.W. led the study of the amenability of the planets for detailed atmospheric characterization. V.V.G. checked the physical parameters of the star. Ad.B. checked the spectral type of the star and determined its metallicity. B.-O.D. took charge of the dynamical simulations. D.B.G. acquired the SpeX spectra. D.K.S. gathered the HCT observations. S.M.L., A.H.M.J.T., P.M. and D.Q. helped to write the manuscript. A.H.M.J.T. prepared most of the figures.

**Author Information** The authors declare no competing financial interests. Readers are welcome to comment on the online version of the paper. Correspondence and requests for materials should be addressed to M.G. (michael.gillon@ulg.ac.be).

Table 1 | Properties of the TRAPPIST-1 planetary system

| Parameter | Value | | |
|---|---|---|---|
| **Star** | **TRAPPIST-1 = 2MASS J23062928-0502285** | | |
| Magnitudes | $V$ = 18.80 ± 0.08, $R$ = 16.47 ± 0.07, $I$ = 14.0 ± 0.1, $J$ = 11.35 ± 0.02, $K$ = 10.30 ± 0.02 | | |
| Distance, $d_\star$ | 12.1 ± 0.4 parsecs (ref. 12) | | |
| Luminosity, $L_\star$ | 0.000525 ± 0.000036 $L_\odot$ (ref. 13) | | |
| Mass, $M_\star$ | 0.080 ± 0.009 $M_\odot$ | | |
| Radius, $R_\star$ | 0.117 ± 0.004 $R_\odot$ | | |
| Density, $\rho_\star$ | $50.3^{+5.7}_{-3.3}\ \rho_\odot$ | | |
| Effective temperature, $T_{eff}$ | 2,550 ± 55 K | | |
| Metallicity, [Fe/H] | +0.04 ± 0.08 (from near-infrared spectroscopy) | | |
| Rotation period, $P_{rot}$ | 1.40 ± 0.05 days (from TRAPPIST photometry) | | |
| Age, $\tau_\star$ | >500 Myr (ref. 13) | | |
| **Planets** | **TRAPPIST-1b** | **TRAPPIST-1c** | **TRAPPIST-1d** |
| Orbital period, $P$ | 1.510848 ± 0.000019 days | 2.421848 ± 0.000028 days | 4.551, 5.200, 8.090, 9.101, 10.401, 12.135, 14.561, **18.202**, 24.270, 36.408, 72.820 days* |
| Mid-transit time, $t_0$ − 2,450,000 ($BJD_{TDB}$) | 7,322.51765 ± 0.00025 | 7,362.72618 ± 0.00033 | 7294.7741 ± 0.0013[†] |
| Transit depth $(R_p/R_\star)^2$ | 0.754 ± 0.025% | 0.672 ± 0.042% | 0.826 ± 0.073%[†] |
| Transit impact parameter $b$ | 0.21 ± 0.14 $R_\star$ | 0.25 ± 0.15 $R_\star$ | 0.24 ± 0.15 $R_\star$[†] |
| Transit duration, $W$ | 36.12 ± 0.46 min | 41.78 ± 0.81 min | 83.3 ± 2.5 min[†] |
| Orbital inclination, $i$ | 89.41 ± 0.41 deg | 89.50 ± 0.31 deg | 89.87 ± 0.10 deg[†] |
| Orbital eccentricity, $e$ | 0 (fixed) | 0 (fixed) | 0 (fixed) |
| Radius, $R_p$ | 1.113 ± 0.044 $R_{Earth}$ | 1.049 ± 0.050 $R_{Earth}$ | 1.168 ± 0.068 $R_{Earth}$[†] |
| Scale parameter, $a/R_\star$ | $20.45^{+0.43}_{-0.81}$ | $28.0^{+0.6}_{-1.1}$ | 41–271[‡] |
| Semi-major axis, $a$ | 0.01111 ± 000040 AU | 0.01522 ± 0.00055 AU | 0.022–0.146 AU[‡] |
| Irradiation, $S_p$ | 4.25 ± 0.38 $S_{Earth}$ | 2.26 ± 0.21 $S_{Earth}$ | 0.02–1.0 $S_{Earth}$[‡] |
| Equilibrium temperature, $T_{eq}$ | | | |
| with Bond albedo of 0.00 | 400 ± 9 K | 342 ± 8 K | 110–280 K[‡] |
| with Bond albedo of 0.75 | 285 ± 7 K | 242 ± 6 K | 75–200 K[‡] |

The values and 1$\sigma$ errors given for the planetary parameters and for the stellar mass ($M_\star$), radius ($R_\star$), density ($\rho_\star$), and effective temperature ($T_{eff}$) were deduced from a global analysis of the photometric data, including *a priori* knowledge of the stellar properties (see Methods). $BJD_{TDB}$, barycentric Julian date in the barycentric Julian time standard. $L_{\odot}$, $M_{\odot}$, $R_{\odot}$ and $\rho_{\odot}$ are, respectively, the luminosity, mass, radius and density of the Sun. $R_p$ and $S_p$ are, respectively, the radius and irradiation of the planet.

*These are the potential orbital periods of TRAPPIST-1d, derived from non-continuous observations. The value in bold type is the most likely value for the period, as derived from the shape of the transits.

†Values calculated on the basis that $P$ = 18.20175 ± 0.00045 days.

‡The ranges allowed by the set of possible periods.

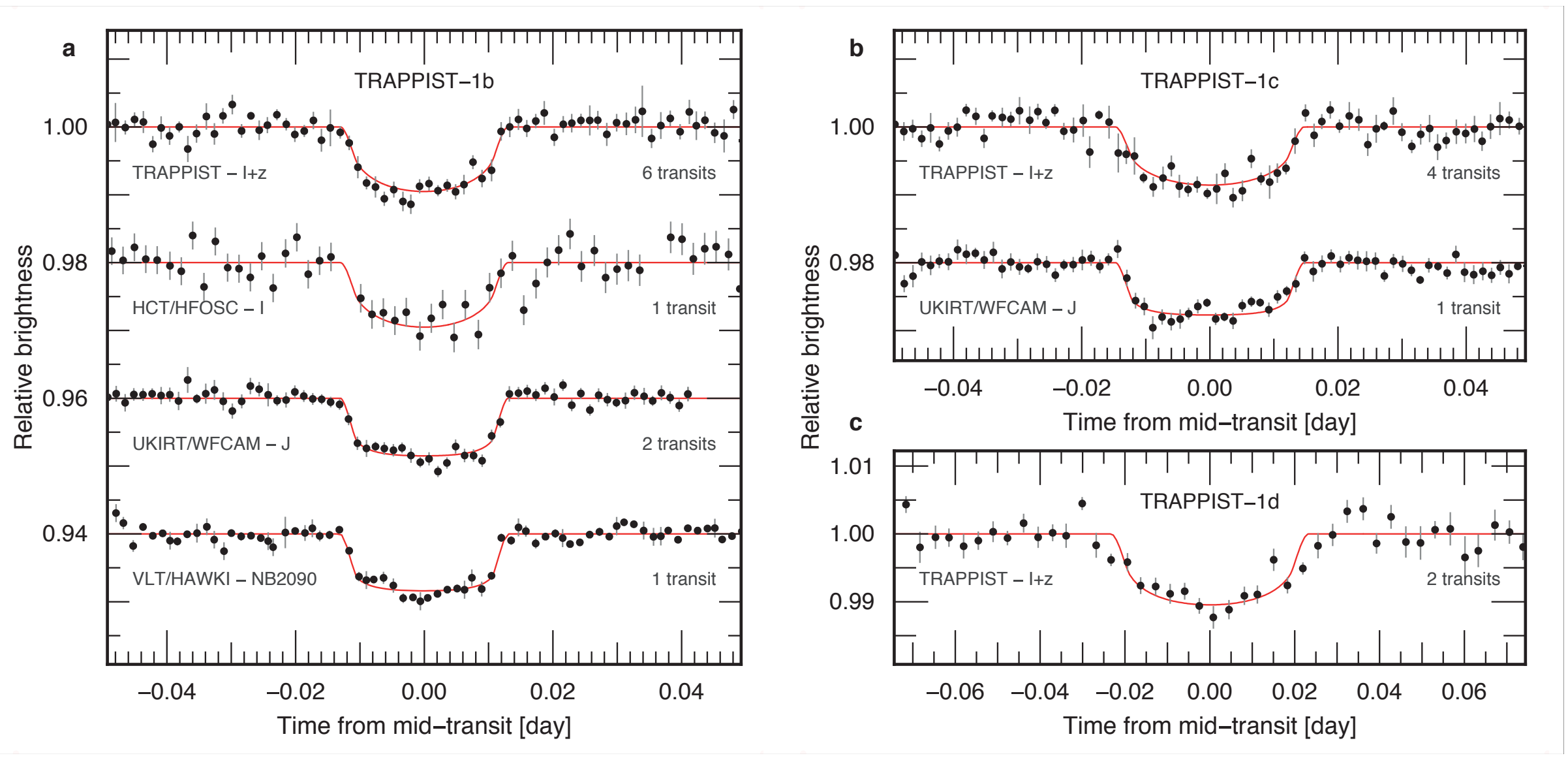

**Figure 1 | Transit photometry of the TRAPPIST-1 planets.** Each light curve is phased to the time of inferior conjunction (mid-transit time) of the object. The light curves are binned in two-minute intervals for planet TRAPPIST-1b (**a**), and in five-minute intervals for planets TRAPPIST-1c (**b**) and TRAPPIST-1d (**c**). The best-fit transit models, as derived from a global analysis of the data, are overplotted (red lines). The light curves are shifted along the *y*-axis for the sake of clarity. For the HCT/Hanle faint object spectrograph camera (HFOSC) light curve, the data are unbinned and the error bars are the formal measurement errors. For the other light curves, the error bars are the standard errors of the mean of the measurements in the bin. WFCAM, wide-field infrared camera on the UKIRT; HAWK-I, high acuity wide field K-band imager on the VLT.

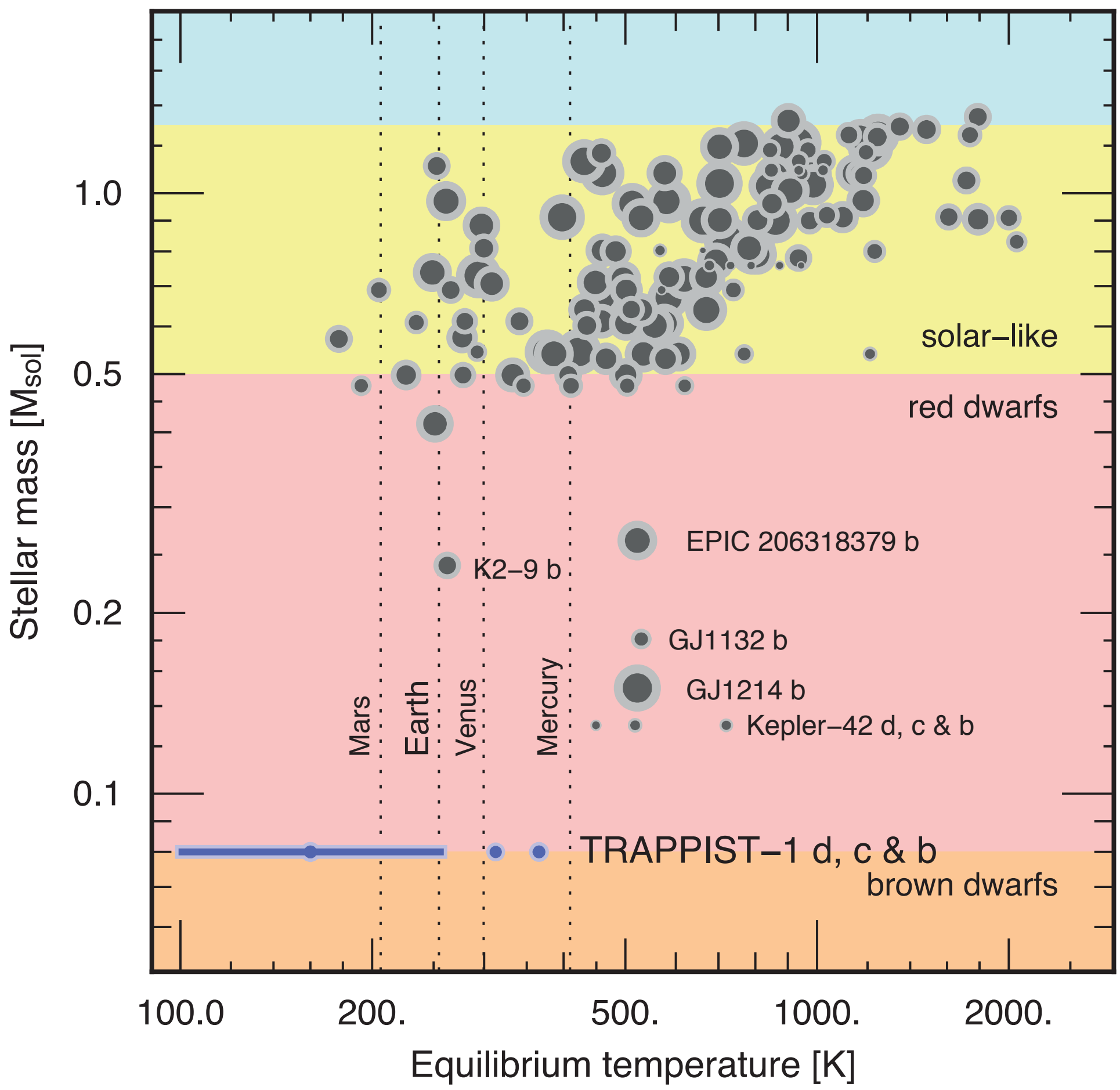

**Figure 2 | Masses of the host stars and equilibrium temperatures of known sub-Neptune-sized exoplanets.** The size of the symbols scales linearly with the radius of the planet. The background is colour-coded according to stellar mass (in units of the Sun's mass). The TRAPPIST-1 planets are at the boundary between planets associated with hydrogen-burning stars and planets associated with brown dwarfs. Equilibrium temperatures are estimated neglecting atmospheric effects and assuming an Earth-like albedo of 0.3. The positions of the Solar System terrestrial planets are shown for reference. The range of possible equilibrium temperatures of TRAPPIST-1d is represented by a solid bar; the dot indicates the most likely temperature. Only the exoplanets with a measured radius equal to or smaller than that of GJ1214b are included.

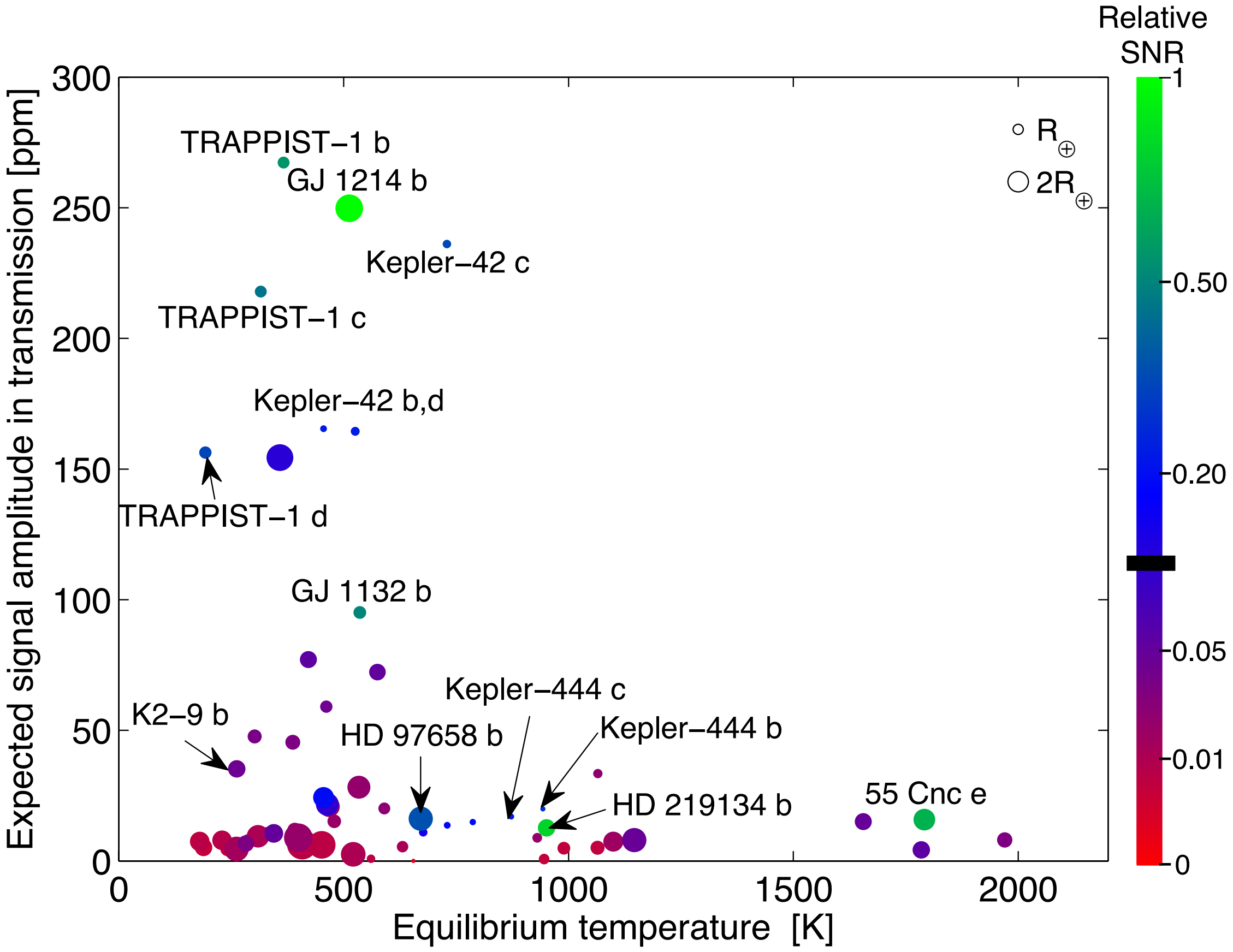

**Figure 3 | Potential for characterizing the atmospheres of known transiting sub-Neptune-sized exoplanets.** The signal being transmitted from each planet is estimated in parts per million (p.p.m.) and for transparent water-dominated atmospheres with a mean molecular weight, $\mu$, of 19. The signal-to-noise ratio (SNR) in transmission (normalized to that of GJ1214b under the same atmospheric assumptions) is also calculated. The estimated signal and SNR are plotted against equilibrium temperatures, assuming a Bond albedo of 0.3. The black horizontal line indicates the SNR that will require 200 (or 500) [or 1,000] hours of in-transit observations with the James Webb Space Telescope to yield a planet's atmospheric temperature with a relative uncertainty below 15% and with abundances within a factor of four in the case of a $H_2O$ (or $N_2$) [or $CO_2$]-dominated atmosphere ($\mu$ = 19 (or 28) [or 39]). Only the exoplanets with a measured radius equal to or smaller than that of GJ1214b are included in the figure. The size of the circular symbol for each planet is proportional to the planet's physical size. For illustration, symbols for planets of one ($R_\oplus$) and two Earth-radii (2 $R_\oplus$) are shown at the top right of the graph.

## METHODS

### Spectral type, parallax and age of the star

TRAPPIST-1 = 2MASS J23062928-0502285 was discovered in 2000 by a search for nearby ultracool dwarfs according to photometric criteria[31], and identified as a high proper-motion (right ascension component $\mu_\alpha = 0.89''$, declination component $\mu_\delta = -0.42''$), moderately active (logarithm of $H_\alpha$ to bolometric luminosity ratio $\log L_{H\alpha}/L_{bol} = -4.61$), M7.5 dwarf at approximately 11 parsecs from Earth. Subsequent findings converged on a spectral type of M8.0 ± 0.5 (refs 11,32), while confirming a moderate level of activity typical of stars of similar spectral type in the vicinity of the Sun[15,33,34]. The spectral classification was checked by comparing a low-resolution ($R$ ~150) near-infrared spectrum of the star[13]—obtained with the SpeX spectrograph[35] mounted on the 3-metre NASA Infrared Telescope Facility—with several spectral-type standards; the spectrum of TRAPPIST-1 best fit that of the M8-type standard LHS 132 (Extended Data Fig. 3a). The Cerro Tololo Interamerican Observatory Parallax Investigation (CTIOPI) project reported the star's trigonometric parallax to be $\pi = 82.6 \pm 2.6$ mas (ref. 12), which translates to a distance of $12.1 \pm 0.4$ parsecs. High-resolution optical spectroscopy failed to detect significant absorption at the 6,708 Å lithium line[36], suggesting that the object is not a very young brown dwarf, but rather a very-low-mass main-sequence star. This is in agreement with its thick disk kinematics[36], its relatively slow rotation (projected rotational velocity $v \sin i = 6 \pm 2$ km s$^{-1}$)[15], its moderate activity, and its reported photometric stability[16], all of which point to an age of at least 500 Myr (ref. 13).

### Metallicity of the star

We obtained new, near-infrared (0.9–2.5 μm) spectrographic data for TRAPPIST-1 with the SpeX spectrograph on the night of 18 November 2015 (universal time), during clear conditions and 0.8″ seeing at K-band. We used the cross-dispersed mode and 0.3″ × 15″ slit, aligned at the parallactic angle, to acquire moderate-resolution data ($\lambda/\Delta\lambda \approx 2{,}000$) with a dispersion of 3.6 Å per pixel, covering the spectral range 0.9–2.5 μm in seven orders. Ten exposures each of 300 seconds were obtained over an air mass ranging from 1.14 to 1.17, followed by observations of the A0V star 67 Aqr (V = 6.41) at an air mass of 1.19 for telluric and flux calibration, as well as internal lamp exposures. Data were reduced using the SpeXtool package version 4.04 (refs 37,38). The reduced spectrum has a median signal-to-noise ratio of 300 in the 2.17–2.35-μm region (see Extended Data Fig. 3b; the metallicity-sensitive atomic features of Na I (2.206 μm, 2.209 μm) and Ca I (2.261 μm, 2.263 μm, 2.266 μm) are labelled). We measured the equivalent widths of these features and the $H_2O$–K2 index (defined in ref. 39), and used the mid- and late-M-dwarf metallicity calibration of ref. 40 to determine [Fe/H] = 0.04 ± 0.02 (measurement) ± 0.07 (systematic) for TRAPPIST-1. The quadratic sum of the two errors resulted in our final value for [Fe/H] of 0.04 ± 0.08.

**Basic parameters of the star**

A recent study[13] derived a luminosity for TRAPPIST-1 of 0.000525 ± 0.000036 $L_\odot$ (where $L_\odot$ is the luminosity of the Sun), using as input data the trigonometric parallax and VRI magnitudes as measured by the Cerro Tololo Interamerican Observatory Parallax Investigation (CTIOPI) project[12], the 2MASS JHK magnitudes[41], the Wide-Field Infrared Survey Explorer (WISE) W123 magnitudes[42], an optical spectrum measured with the Kitt Peak National Observatory Ritchey–Chretien spectrograph[43], and a near-infrared spectrum measured by SpeX/Prism. Using this luminosity and an age constraint of >500 Myr, the authors of ref. 13 derived (from evolutionary-model isochrones and the Stefan–Boltzmann law) the following values for the mass, radius and effective temperature of TRAPPIST-1: $M_\star = 0.082 \pm 0.009\ M_\odot$, $R_\star = 0.116 \pm 0.004\ R_\odot$, and $T_{\rm eff} = 2557 \pm 64$ K. To account for the uncertainties coming from the assumptions and details of the evolutionary models, we carried out a new determination of these three basic parameters, using recent solar metallicity evolutionary-model isochrones that consistently couple atmosphere and interior structures[44]. We obtained $M_\star = 0.089\ M_\odot$, $R_\star = 0.112\ R_\odot$, and $T_{\rm eff} = 2{,}615$ K. We then added the difference between the two determinations quadratically to the errors of ref. 13, adopting finally $M_\star = 0.082 \pm 0.011\ M_\odot$, $R_\star = 0.116 \pm 0.006\ R_\odot$, and $T_{\rm eff} = 2{,}555 \pm 85$ K. We took the normal distributions corresponding to these values and errors as prior probability distribution functions in the Bayesian analysis of our photometric data (see below).

**Possible binary nature of the star**

High-resolution imaging from the ground[45-47] and from space with the Hubble Space Telescope[48] discarded the existence of a companion down to an angular distance of 0.1″, corresponding to a projected physical distance of 1.2 AU at 12 parsecs, and in good agreement with the reported stability of the radial velocity of the star at the ~10 $\mathrm{ms}^{-1}$ level over one week[49] and at the ~150 $\mathrm{ms}^{-1}$ level over about ten weeks[50]. We performed spectral binary template fitting[51] to the IRTF/SpeX spectroscopy, and statistically reject the presence of an L- or T-type brown-dwarf companion that would be visible in a blended-light spectrum. TRAPPIST-1 can thus, in all probability, be considered to be an isolated star.

**Upper magnitude limits on a background eclipsing binary**

We measured the J2000 equatorial coordinates of TRAPPIST-1 in the 2015 TRAPPIST images, using 29 stars from the UCAC2 catalogue[52] and the Pulkovo Observatory Izmccd astrometric software[53]. We obtained coordinates of right ascension (RA) = 23 h 06 min 30.34 s and declination (Dec.) = −05°02′ 36.44″. Owing to the high proper motion of TRAPPIST-1 (~1″ per year), we could assess the possible presence of a background object by examining this exact position in several previous images taken from the POSS (1953; ref. 54) and 2MASS (1998; ref. 41) image catalogues. We detected no possible additional source at this position in any of these images. The faintest stars detected at other positions in the 2MASS images have J-band magnitudes of ~17. We adopt this value as an absolute lower threshold for the J-band magnitude of a background source blended with TRAPPIST-1 in our TRAPPIST 2015 images. TRAPPIST-1 has a J-band magnitude of 11.35 (ref. 42), and the achromaticity of the transits of TRAPPIST-1b as observed from 0.85 μm to 2.09 μm means that, if the transits originated from a background eclipsing binary (BEB), then that BEB would have to be a very red object with a spectral type similar to that of TRAPPIST-1. Combining these two facts, the BEB scenario would require an unphysical eclipse depth of more than 100% in the photometric bands probed by our observations to match the ~0.8% depths measured after dilution by the light from TRAPPIST-1. We thus firmly discard the BEB scenario.

**Photometric observations and analysis**

The TRAPPIST[8,55] observations in which the transits were detected consisted of 12,295 exposures, each of 55 seconds, gathered with a thermoelectrically cooled 2Kx2K CCD camera (field of view of 22′ × 22′; pixel scale of 0.65″). Most of the observations were obtained through an I+z filter with a transmittance greater than 90% from 750 nm to beyond 1,100 nm—the effective bandpass in this spectral range being defined by the response of the CCD. On the basis of the spectral efficiency model for TRAPPIST and an optical spectrum of a spectroscopic standard M8V star (VB10), we compute an effective wavelength of 885 ± 5 nm for these observations. For the nights of 20 November and 19 December 2015, the target was close to the full Moon and the observations were performed in the Sloan z′ filter to minimize the background. After a standard pre-reduction (bias, dark, flat-field correction), the TRAPPIST automatic pipeline extracted the stellar fluxes from the images using the DAOPHOT aperture photometry software[56] for eight different apertures. A careful selection of both the photometric aperture size and the stable comparison stars was then performed manually to obtain the most accurate differential light curves of the target.

Photometric follow-up observations were performed with the HAWK-I near-infrared imager[57] on the European Southern Observatory (ESO) 8-metre Very Large Telescope (Chile), with the HFOSC optical spectro-imager[58] on the 2-metre Himalayan Chandra Telescope (India), and with WFCAM[59] located at the prime focus of the 3.8-metre UKIRT telescope (Hawaii).

The VLT/HAWK-I observations of a transit of planet TRAPPIST-1b were performed during the night of 8 November 2015. HAWK-I is composed of four Hawaii 2RG 2,048 × 2,048 pixel detectors (pixel scale = 0.106″). Its total field of view on the sky is 7.5′ × 7.5′. The transit was observed through the narrowband filter NB2090 ($\lambda = 2.095$ μm, width = 0.020 μm). 185 exposures, composed of 17 integrations of 1.7 seconds each, were acquired during the run in 'stare' mode—that is, without applying a jitter pattern. After standard calibration of the images, stellar flux measurement was performed by aperture photometry[14].

The HCT/HFOSC observations of a transit of TRAPPIST-1b were performed on 18 November 2015. The imager in the HFOSC CCD detector is an array of 2,048 × 2,048 pixels, corresponding to a field of view of 10′ × 10′ on-sky (pixel scale = 0.3″). The observations consisted of 104 exposures, each of 20 seconds, taken in stare mode and in the I filter, centred on the expected transit time. After a standard calibration of these images and their photometric reduction with DAOPHOT, differential photometry was performed. We estimate the effective wavelength of these observations to be 840 ± 20 nm, given the spectral response of HFOSC and an optical spectrum of the M8V standard star VB10.

The UKIRT/WFCAM observations of two transits of planet TRAPPIST-1b and one transit of planet TRAPPIST-1c consisted of three runs of 4 hours each, performed on 5, 6 and 8 December 2015 in the J-band. WFCAM is composed of four HgCdTe detectors of 2,048 × 2,048 pixels each, with a pixel scale of 0.4″, resulting in a field of view of 13.65′ × 13.65′ for each detector. On 5 December 2015, 1,365 exposures composed of three integrations of 2 seconds each were performed in stare mode. For the runs on 6 and 8 December 2015, respectively, 1,181 and 1,142 exposures composed of five one-second exposures were performed, again in stare mode and using the same pointing as on 5 December 2015. Differential aperture photometry was performed with DAOPHOT on all calibrated images.

**Global analysis of the photometry**

We inferred the parameters of the three detected planets transiting TRAPPIST-1 from analysis of their transit light curves (Extended Data Fig. 1 and Extended Data Table 1) with an adaptive Markov-chain Monte Carlo (MCMC) code[14]. We converted each universal time (UT) of mid-exposure to the $BJD_{TDB}$ time system[60]. The model assumed for each light curve was composed of the eclipse model of ref. 61, multiplied by a baseline model, aiming to represent the other astrophysical and instrumental mechanisms able to produce photometric variations. Assuming the same baseline model for all light curves, and minimizing the Bayesian information criterion (BIC)[62], we selected a second-order time polynomial as a baseline model to represent the curvature of the light curves due to the differential extinction and the low-frequency variability of the star, and added an instrumental model composed of a second-order polynomial function of the positions and widths of the stellar images.

Stellar metallicity, effective temperature, mass and radius were four free parameters in the MCMC for which prior probability distribution functions (PDFs) were selected as input. Here, the normal distributions $N(0.04, 0.08^2)$ dex, $N(2{,}555, 85^2)$ K, $N(0.082, 0.011^2)$ $M_\odot$, and $N(0.116, 0.006^2)$ $R_\odot$ were assumed on the basis of *a priori* knowledge of the stellar properties (see the section on 'Basic parameters of the star'). Circular orbits were assumed for all transiting objects. For each of them, the additional free parameters in the MCMC included: (1) the transit depth $dF$, defined as $(R_p/R_\star)^2$, with $R_p$ and $R_\star$ being the planetary and stellar radii, respectively; (2) the transit impact parameter $b = a \cos i/R_{\star}$, with $a$ and $i$ being the planet's semi-major axis and orbital inclination, respectively; (3) the orbital period $P$; (4) the transit width $W$ defined as $(P \times R_{\star}/a)\ [(1+R_p/R_{\star})^2 - b^2]^{1/2}/\pi$ ; and (5) the mid-transit time (time of inferior conjunction) $T_0$. Uniform prior distributions were assumed for each of these free parameters. At each step of the MCMC, values for $R_p$, $a$ and $i$, were computed from the values for the transit and stellar parameters; values were also computed for the irradiation of the planet in Earth units and for its equilibrium temperatures, assuming Bond albedos of 0 and 0.75, respectively. A quadratic limb-darkening law[62] was assumed for the star. For each bandpass, values and errors for the limb-darkening coefficients $u_1$ and $u_2$ were derived from the tables in ref. 63 (Extended Data Table 2), and the corresponding normal distributions were used as prior PDFs in the MCMC. $u_1$ and $u_2$ were free parameters under the control of these PDFs in the MCMC.

We divided our analysis into three phases. The first phase focused on the two inner planets, for which the period is firmly determined. A circular orbit was assumed for both planets. All transit light curves of the two planets were used as input data for this first phase, except the TRAPPIST light curve of 11 December 2015, for which the transit of planet TRAPPIST-1c is blended with a transit of planet TRAPPIST-1d. A preliminary MCMC analysis composed of one chain of 50,000 steps was first performed to estimate the need to rescale the photometric errors[14]. Then a longer MCMC analysis was performed, composed of five chains of 100,000 steps, whose convergence was checked using the statistical test of ref. 64. The parameters derived from this analysis for the star and its two inner planets are shown in Table 1. We performed a similar analysis assuming a uniform prior PDF for the stellar radius to derive the value of the stellar density constrained only by the transit photometry[65]. It resulted in a stellar density of $49.3^{+4.1}_{-8.3}\ \rho_{\odot}$, in excellent agreement with the density of $55.3 \pm 12.1\ \rho_\odot$ derived from the *a priori* knowledge of the star, thus bringing a further validation of the planetary origin of the transit signals.

In the second phase of our analysis, we performed 11 global MCMC analyses of all transit light curves, each of them consisting of one chain of 50,000 steps and corresponding to one of the possible values of the period of TRAPPIST-1d (see Table 1) for which a circular orbit was assumed. We then repeated the 11 analyses under the assumption of an eccentric orbit for TRAPPIST-1d. We used the medians of the BIC posterior distributions to compare the relative posterior probability of each orbital model through the formula $P1/P2 = e^{(BIC_2 - BIC_1)/2}$. The resulting relative probabilities are given in Extended Data Table 3. The table shows that our data favour (with a relative probability of >10%) a circular orbit and an orbital period of between 10.4 and 36.4 days—the most likely period being 18.4 days.

In the final phase, we performed individual analyses of the light curves to measure the mid-eclipse time of each transit to support future TTV studies of the system[22,66]. The resulting timings are shown in Extended Data Table 4. They do not reveal any significant TTV signal, which is not surprising given the amplitude of the expected periodicity departures (see below) combined with the limited timing precision of the TRAPPIST photometry.

Extended Data Figs 1 and 2 show the raw and de-trended light curves, respectively; for each of these, the best-fit eclipse plus baseline model is overplotted. The phased-time de-trended light curves are shown for each planet and bandpass in Fig. 1.

#### Photometric variability of the star

We used the TRAPPIST data set to assess the photometric variability of the star at about 900 nm. On a timescale of a few hours—corresponding to the typical duration of our observing runs—the star appears to be relatively stable, except for the transits and for four sharp, low-amplitude increases in brightness (of one to a few per cent) that are followed by exponential-type decreases to normal levels within 10−15 minutes (Extended Data Fig. 4), which we attribute to flares[67]. The low amplitude and inferred low frequency (1/60 $h^{-1}$) of these flares is consistent with the reported low level of activity of the star[15,33,34], strengthening the inference that the system is not young.

To assess the lower-frequency variability of TRAPPIST-1, we built its global differential light curve in the I+z filter, using four stable stars of similar brightness in the TRAPPIST images as comparison stars. We filtered out the flares, transits, and measurements taken in cloudy conditions to create the resulting light curve, consisting of 12,081 photometric measurements. Extended Data Fig. 5a compares this light curve to that for the comparison star 2MASS J23063445-0507511. It clearly shows some variability at the level of a few per cent, which is consistent with previous photometric results obtained in the I-band[16]. A Lomb–Scargle (LS) periodogram[68] analysis of the light curve—from which low-frequency variations and differential extinction have been filtered out by division of the best-fit fourth-order polynomial in time and air mass—reveals a power excess with a period of 1.4 days (see Extended Data Fig. 5b). Cutting the light curves in two, and in four in a second test, and performing a LS analysis of each fraction revealed a power excess at about 1.4 days for all of them, supporting a genuine periodic signal of astrophysical origin. Associating it with the stellar rotation period, the resulting equatorial rotation speed of 4.2 km $s^{-1}$ (assuming $R_\star = 0.117\,R_\odot$) is consistent with the literature measurement[15] for $v$ sin$i$ of $6 \pm 2$ km $s^{-1}$, making this association physically meaningful. Given the scatter of the peak values obtained in the LS analyses of the light-curve fractions, we estimate the error bar on the rotation period of 1.40 days to be 0.05 days. In summary, the photometric variability of the star seems to be dominated by the rotation and evolution of photospheric inhomogeneities (spots) combined with rare flares.

**Dynamics of the system**

We computed the tidal circularization timescales[69] of the three planets according to $t_{\mathrm{circ}} = \frac{2PQ}{63\pi} \times \frac{M_{\mathrm{p}}}{M_{*}} \times \left( \frac{a}{R_{\mathrm{p}}} \right)^{5}$, assuming planetary masses, $M_{\mathrm{p}}$, ranging from 0.45 Earth masses (pure ice composition) to 3 Earth masses (pure iron composition)[22] and a tidal quality factor[70], $Q$, of 100, corresponding to the maximum value derived for terrestrial planets and satellites of the solar system[70]. For planets TRAPPIST-1b and -1c, the computed values range from 22 Myr to 145 Myr and from 177 Myr to 1.1 Gyr, respectively. Taking into account that the system is apparently not very young and that the orbits have weak mutual perturbations (as they are not close to any mean-motion resonance), our assumption of circular orbits for the two inner planets is reasonable. On the other hand, the same computations result in values ranging from a few to tens of billions of years for TRAPPIST-1d, making a significant orbital eccentricity possible from a tidal theory perspective. Nonetheless, a nearly circular orbit for this outer planet is still a reasonable hypothesis when considering the strong anticorrelation of orbital eccentricity and multiplicity of planets detected by radial velocities[71], and is favoured by our global analysis of the transit photometry (see above).

We used the Mercury software package[72] to assess the dynamical stability of the system over 10,000 years for all possible periods of TRAPPIST-1d. Instabilities appeared in our simulations only for the unlikely scenarios of this planet on a significantly eccentric ($e \geq 0.4$) 4.5-day or 5.2-day orbit.

To assess the potential of the TTV method[24,66] to measure the masses of the planets, we integrated the dynamical evolution of the system at high sampling over two years, assuming Earth masses for the three planets and an 18.4-day circular orbit for TRAPPIST-1d. These simulations resulted in TTV amplitudes of several tens of seconds, and led us to conclude that, with an intensive transit monitoring campaign—with instruments able to reach timing precisions of a few tens of seconds (for example, with VLT/HAWK-I or UKIRT/WFCAM; Extended Data Table 4)—it should be possible to constrain the planetary masses.

**Planets' suitability for atmospheric characterization**

We estimated the typical signal amplitude in transit transmission spectroscopy for all the transiting exoplanets with a size equal to or smaller than that of the mini-Neptune GJ1214b (ref. 73). We computed this amplitude as $2R_p h_{eff}/R_\star^2$, where $R_p$ is the planetary radius, $h_{eff}$ is the effective atmospheric height (that is, the extent of the atmospheric annulus), and $R_\star$ is the stellar radius. The effective atmospheric height is directly proportional to the atmospheric scale height, $H = kT/\mu g$, where k is Boltzmann's constant, $T$ is the atmospheric temperature, $\mu$ is the atmospheric mean molecular mass, and $g$ is the surface gravity. The ratio $h_{eff}/H$ for a transparent atmosphere[24,74] is typically between 6 and 10, and thus depends strongly on the presence of clouds and the spectral resolution and range covered. Our estimates (Fig. 2) are based on an $h_{eff}/H$ ratio of 7 and the conservative assumption of a volatile-dominated atmosphere ($\mu = 20$) with a Bond albedo of 0.3. All other parameters for the planets were derived from exoplanets.org[75]. As an illustration, the maximum transit depth variations projected under those assumptions for GJ1214b are about 250 p.p.m., in agreement with independent simulations[76].

For the same sample of planets, we also derived the typical SNRs in transit transmission spectroscopy from the ratio of our computed signal amplitudes over the square root of the flux (determined from the J-band magnitudes of the host stars). The SNRs of TRAPPIST-1's planets in transmission are expected to range between 0.22 and 0.55 times the one of GJ 1214b under the same theoretical assumptions, suggesting that these planets are well suited for atmospheric studies with HST/WFC3 similar to those previously targeting GJ1214b (refs 76,77).

Given published simulations for terrestrial planets[24], we estimate that characterization of TRAPPIST-1b, -1c and -1d should require up to 70 hours, 90 hours and 270 hours, respectively, of in-transit observations with the James Webb Space Telescope (JWST), and should yield atmospheric temperatures with relative uncertainties below 15% and abundances within a factor of four. Assuming that the atmospheres of TRAPPIST-1's planets are not depleted and do not harbour a high-altitude cloud deck, JWST should, notably, yield constraints on the abundances of molecules with large absorption bands such as $H_2O$, $CO_2$, $CH_4$, CO and $O_3$ if their abundances are at or greater than the 10-p.p.m. level.

We also assessed the potential of the cross-correlation technique[78] to constrain the atmospheric properties of the TRAPPIST-1 planets, following a published formalism[79]. We find that detecting $O_2$ in TRAPPIST-1's planets should require up to 80 transit observations with one of the next-generation, giant ground-based telescopes. Taking into account the limited fraction of transits visible at low air mass, such an endeavour could be reached in 5 to 15 years.

**Code availability**

Equivalent widths and $H_2O$–K2 index measurements in the SpeX spectra were made using the IDL program created by A. Mann and distributed at http://github.com/awmann/metal. Conversion of the UT times for the photometric measurements to the $BJD_{TDB}$ system was performed using the online program created by J. Eastman and distributed at http://astroutils.astronomy.ohio-state.edu/time/utc2bjd.html. The Image Reduction and Analysis Facility (IRAF) software is distributed by the National Optical Astronomy Observatory, which is operated by the Association of Universities for Research in Astronomy, Inc., under cooperative agreement with the National Science Foundation. The MCMC software used to analyse the photometric data is a custom Fortran 90 code that can be obtained upon request.

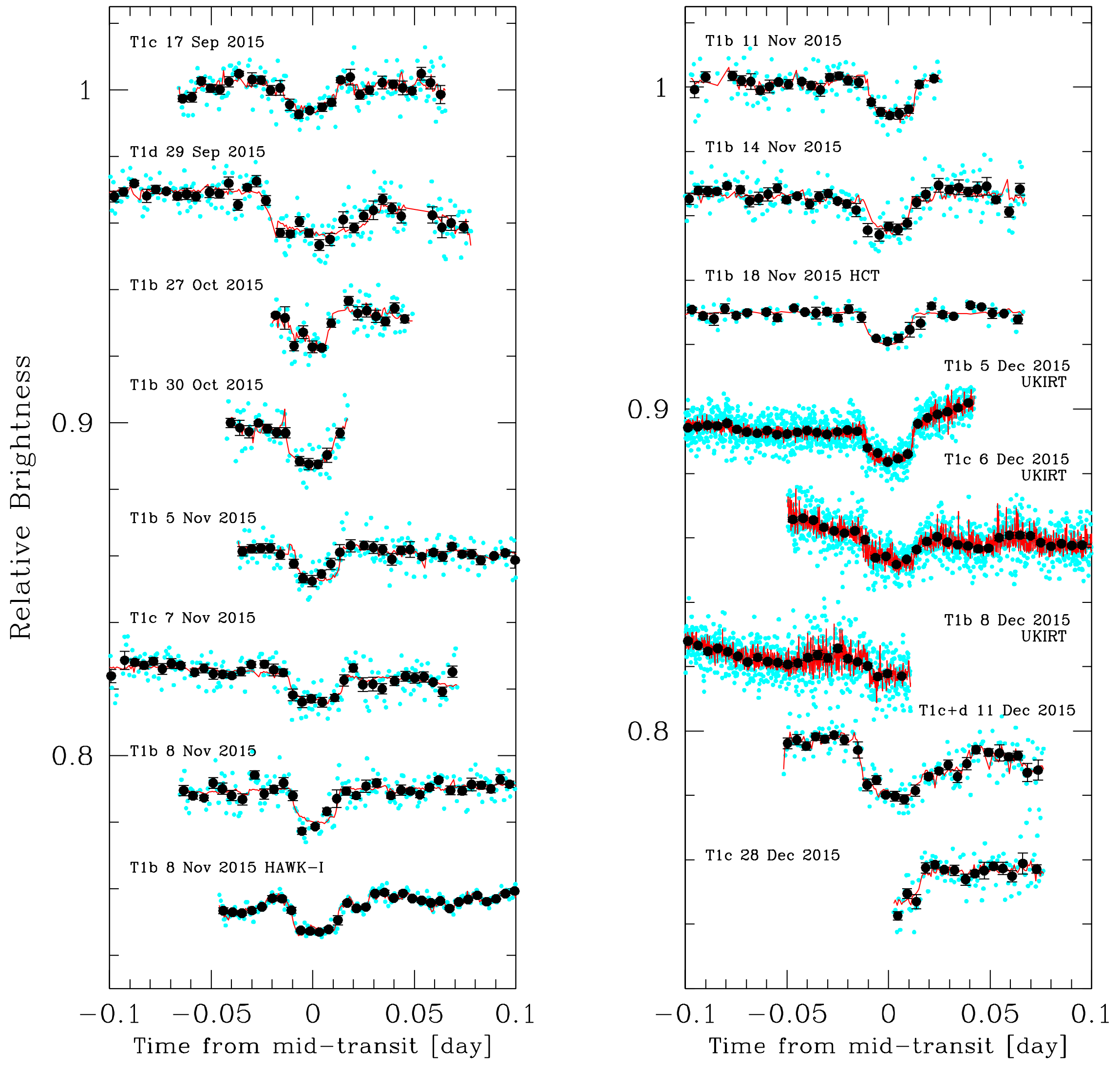

**Extended Data Figure 1 | Raw TRAPPIST-1 transit light curves.** The light curves are shown in chronological order from top to bottom and left to right, with unbinned data shown as cyan dots, and binned 0.005-day (7.2-minute) intervals shown as black dots with error bars. The error bars are the standard errors of the mean of the measurements in the bins. The best-fit transit-plus-baseline models are overplotted (red line). The light curves are phased for the mid-transit time and shifted along the *y*-axis for clarity. For the dual transit of 11 December 2015, the light curve is phased for the mid-transit time of planet TRAPPIST-1c.

–

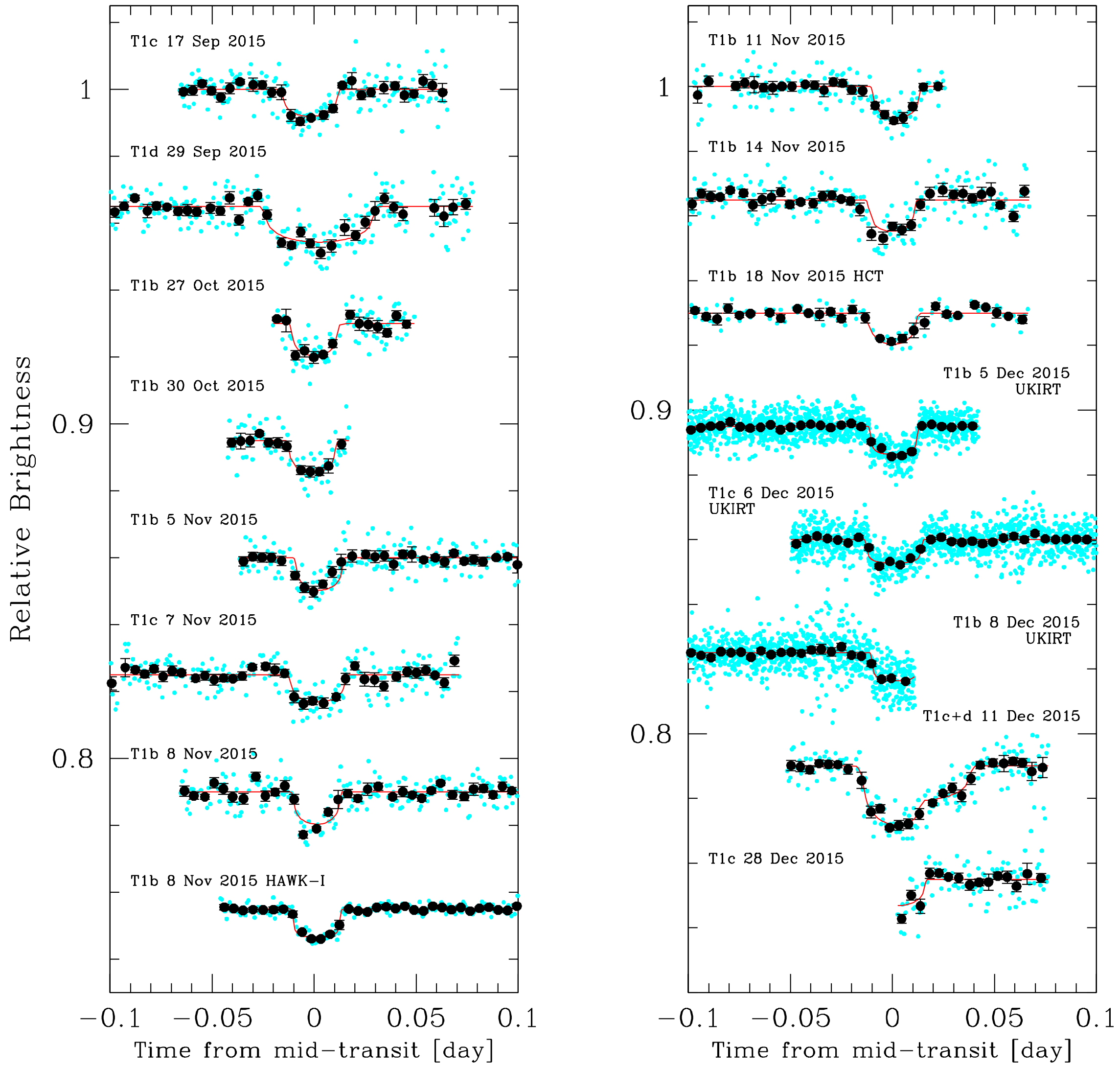

**Extended Data Figure 2 | De-trended TRAPPIST-1 transit light curves.** The details are as in Extended Data Fig. 1, except that the light curves shown here are divided by the best-fit baseline model to highlight the transit signatures.

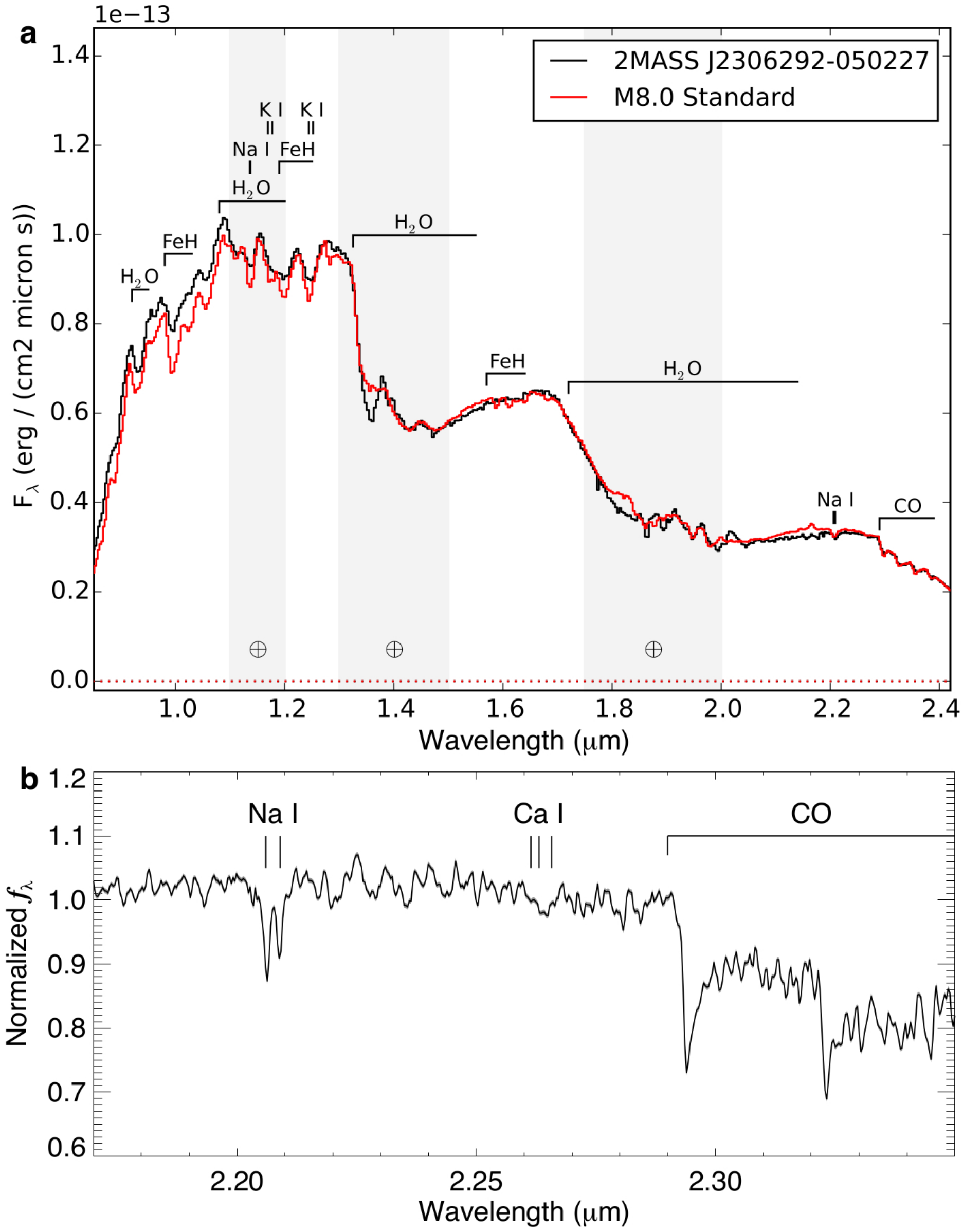

**Extended Data Figure 3 | Near-infrared spectra of TRAPPIST-1.** **a**, Comparison of TRAPPIST-1's near-infrared spectrum (black)—obtained with the spectrograph IRTF/SpeX[35]—with that of the M8-type standard LHS132 (red). **b**, Cross-dispersed IRTF/SpeX spectrum of TRAPPIST-1 in the 2.17–2.35-μm region. Na I, Ca I and CO features are labelled. Additional structure primarily originates from overlapping $H_2O$ bands. The spectrum is normalized at 2.2 μm. $F_\lambda$ ,spectral flux density; $f_\lambda$, normalized spectral flux density.

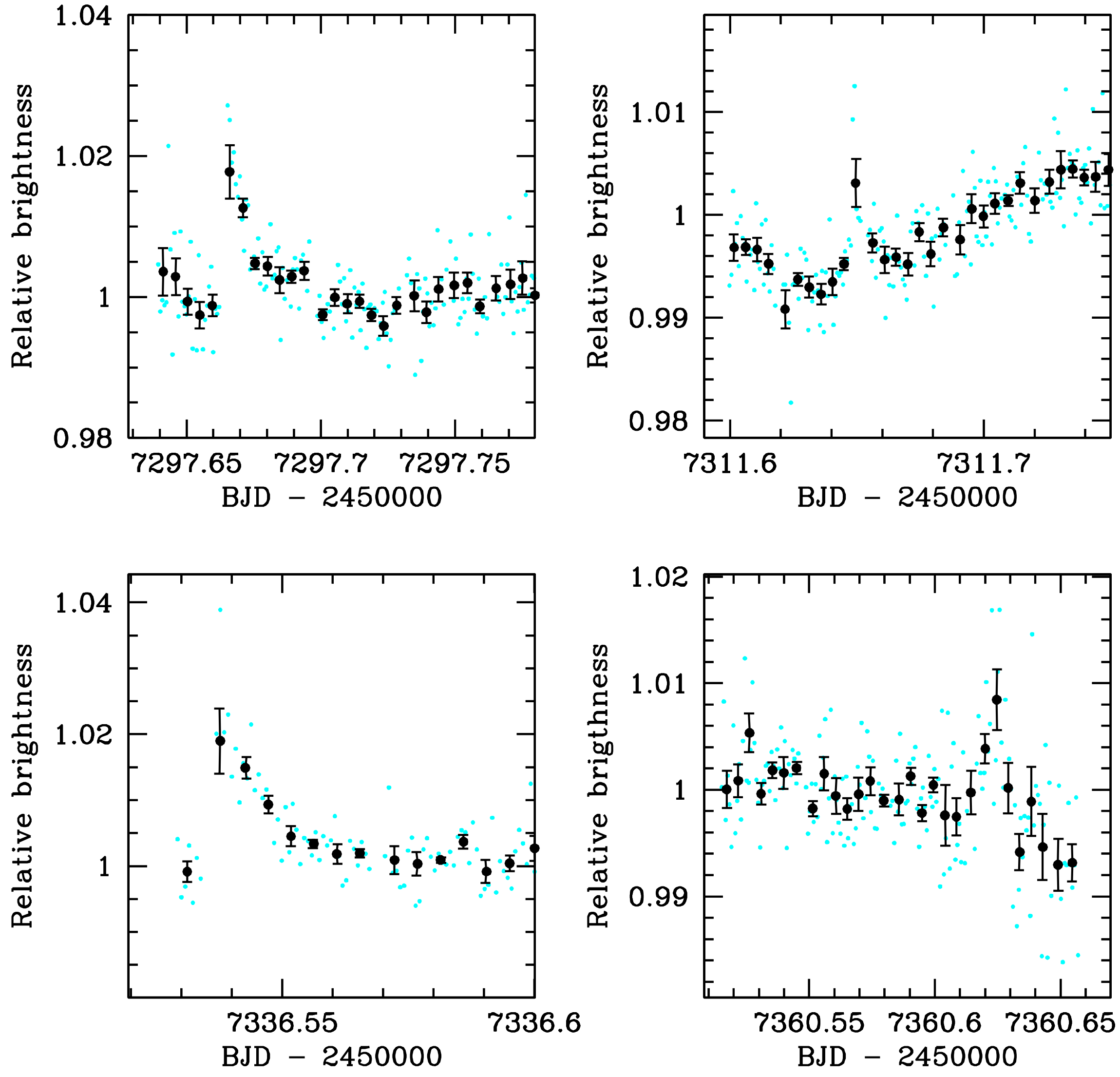

**Extended Data Figure 4 | Flare events in the TRAPPIST 2015 photometry.** The photometric measurements are shown unbinned (cyan dots) and binned per 7.2-minute interval (black dots). For each interval, the error bars are the standard error of the mean.

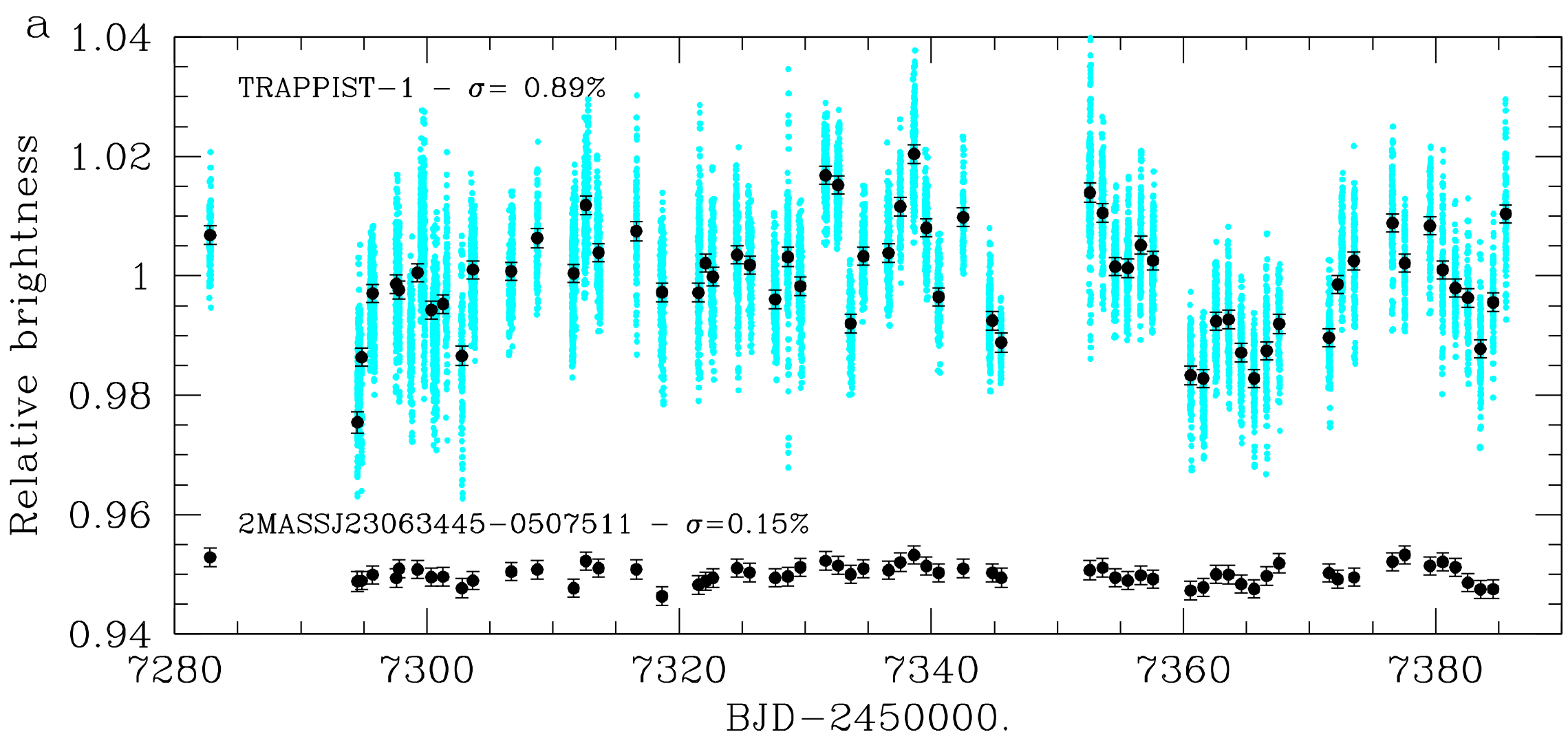

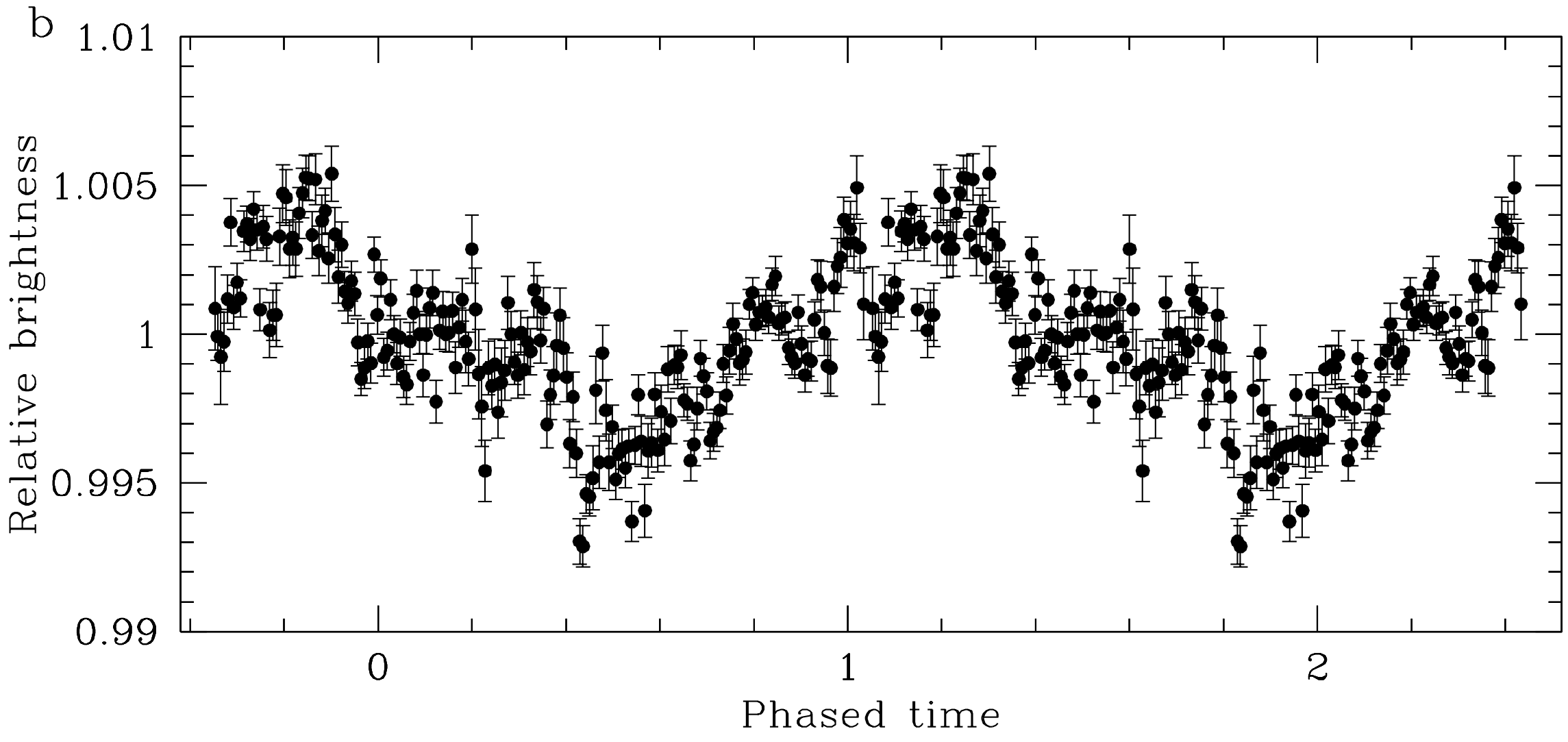

**Extended Data Figure 5 | Photometric variability of TRAPPIST-1. a**, Global light curve of the star as measured by TRAPPIST. The photometric measurements are shown unbinned (cyan dots) and binned per night (black dots with error bars (±s.e.m.)). This light curve is compared with that of the comparison star 2MASS J23063445-0507511, shifted along the *y*-axis for clarity. **b**, The same light curve for TRAPPIST-1, folded on the period $P = 1.40$ days and binned by 10-minute intervals (error bars indicate ±s.e.m.). For clarity, two consecutive periods are shown.

**Extended Data Table 1 | TRAPPIST-1 transit light curves**

| Date | Instrument | Filter | $N_p$ | $T_{exp}$ | Baseline function | Transit(s) |
|---|---|---|---|---|---|---|
| 17 Sep 2015 | TRAPPIST | I+z | 163 | 55s | $p(t^2+xy^2+f^2)$ | TRAPPIST-1c |
| 29 Sep 2015 | TRAPPIST | I+z | 232 | 55s | $p(t^2+xy^2+f^2)$ | TRAPPIST-1d |
| 27 Oct 2015 | TRAPPIST | I+z | 84 | 55s | $p(t^2+xy^2+f^2)$ | TRAPPIST-1b |
| 30 Oct 2015 | TRAPPIST | I+z | 77 | 55s | $p(t^2+xy^2+f^2)$ | TRAPPIST-1b |
| 05 Nov 2015 | TRAPPIST | I+z | 237 | 55s | $p(t^2+xy^2+f^2)$ | TRAPPIST-1b |
| 07 Nov 2015 | TRAPPIST | I+z | 241 | 55s | $p(t^2+xy^2+f^2)$ | TRAPPIST-1c |
| 08 Nov 2015 | TRAPPIST | I+z | 231 | 55s | $p(t^2+xy^2+f^2)$ | TRAPPIST-1b |
| | VLT/HAWK-I | NB2090 | 207 | 17x1.7s | $p(t^2+xy^2+f^2)$ | TRAPPIST-1b |
| 11 Nov 2015 | TRAPPIST | I+z | 140 | 55s | $p(t^2+xy^2+f^2)$ | TRAPPIST-1b |
| 14 Nov 2015 | TRAPPIST | I+z | 241 | 55s | $p(t^2+xy^2+f^2)$ | TRAPPIST-1b |
| 18 Nov 2015 | HCT/HFOSC | I | 103 | 20s | $p(t^2+xy^2+f^2)$ | TRAPPIST-1b |
| 05 Dec 2015 | UKIRT | J | 1312 | 3x2s | $p(t^2+xy^2+f^2)$ | TRAPPIST-1b |
| 06 Dec 2015 | UKIRT | J | 1175 | 5x1s | $p(t^2+xy^2+f^2)$ | TRAPPIST-1c |
| 08 Dec 2015 | UKIRT | J | 1109 | 5x1s | $p(t^2+xy^2+f^2)$ | TRAPPIST-1b |
| 11 Dec 2015 | TRAPPIST | I+z | 158 | 55s | $p(t^2+xy^2+f^2)$ | TRAPPIST-1c + d |
| 28 Dec 2015 | TRAPPIST | I+z | 94 | 55s | $p(t^2+xy^2+f^2)$ | TRAPPIST-1c (partial) |

For each light curve, the date, instrument, filter, number of points ($N_p$), exposure time ($T_{exp}$), and baseline function are given. For the baseline functions, $p(t^2)$, $p(xy^2)$, and $p(f^2)$ denote, respectively, second-order polynomial functions of time, of the $x$ and $y$ positions, and of the full-width at half-maximum of the stellar images.

**Extended Data Table 2 | Quadratic limb-darkening coefficients**

| Bandpass | $u_1$ | $u_2$ |
|---|---|---|
| I (HCT/HFOSC) | 0.72±0.10 | 0.15±0.11 |
| I+z (TRAPPIST) | 0.65±0.10 | 0.28±0.12 |
| J (UKIRT/WFCAM) | 0.10±0.05 | 0.57±0.02 |
| NB2090 (VLT/HAWKI) | 0.04±0.03 | 0.50±0.03 |

We inferred these values and errors for the quadratic coefficients $u_1$ and $u_2$ for TRAPPIST-1 from theoretical tables[63], and used the values and errors as *a priori* knowledge of the stellar limb-darkening in a global MCMC analysis of the transit light curves. The error bars were obtained by propagation of the errors on the stellar gravity, metallicity, and effective temperature.

**Extended Data Table 3 | Posterior likelihoods of the orbital solutions for TRAPPIST-1d**

| TRAPPIST-1d period (d) | Circular orbit | Eccentric orbit | $a$ (au) | $S_p$ ($S_{Earth}$) |
|---|---|---|---|---|
| 4.551 | 0.0016 | 0.0017 | 0.023 | 0.98 |
| 5.200 | 0.0041 | 0.0045 | 0.025 | 0.82 |
| 8.090 | 0.012 | 0.013 | 0.034 | 0.45 |
| 9.101 | 0.018 | 0.011 | 0.037 | 0.39 |
| 10.401 | 0.139 | 0.0067 | 0.040 | 0.33 |
| 12.135 | 0.243 | 0.0029 | 0.045 | 0.26 |
| 14.561 | 0.393 | 0.0023 | 0.050 | 0.21 |
| 18.204 | 1 | 0.0018 | 0.058 | 0.15 |
| 24.270 | 0.212 | 0.0016 | 0.071 | 0.11 |
| 36.408 | 0.122 | 0.0014 | 0.093 | 0.06 |
| 72.820 | $7.5e^{-5}$ | $6.8e^{-8}$ | 0.147 | 0.02 |

The likelihoods shown for the circular and eccentric orbits are normalized to the most likely solution (that is, a circular orbit of $P$ = 18.204 days (d)). For each orbit, the semi-major axis, $a$ (in astronomical units (AU)), assuming a stellar mass of 0.08 $M_\odot$ (Table 1), and the mean irradiation, $S_p$ (in Earth units ($S_{Earth}$)) are shown.

**Extended Data Table 4 | Individual mid-transit timings measured for the TRAPPIST-1 planets**

| Planet | Instrument | Epoch | Mid-transit timing ($BJD_{TDB}$-2,450,000) |
|---|---|---|---|
| TRAPPIST-1b | TRAPPIST | 0 | $7322.5161^{+0.0013}_{-0.0010}$ |
| | TRAPPIST | 2 | $7325.5391^{+0.0035}_{-0.0013}$ |
| | TRAPPIST | 6 | 7331.5803±0.0013 |
| | TRAPPIST | 8 | 7334.6038±0.0012 |
| | VLT/HAWK-I | 8 | 7334.60490±0.00020 |
| | TRAPPIST | 10 | 7337.6249±0.0010 |
| | TRAPPIST | 12 | $7340.6474^{+0.0010}_{-0.0022}$ |
| | HCT/HFOSC | 15 | 7345.18011±0.00089 |
| | UKIRT/WFCAM | 26 | 7361.79960±0.00030 |
| | UKIRT/WFCAM | 28 | 7364.82137±0.00056 |
| TRAPPIST-1c | TRAPPIST | 0 | 7282.8058±0.0010 |
| | TRAPPIST | 21 | 7333.6633±0.0010 |
| | UKIRT/WFCAM | 33 | 7362.72623±0.00040 |
| | TRAPPIST | 35 | 7367.5699±0.0012 |
| | TRAPPIST | 42 | 7384.5230±0.0011 |
| TRAPPIST-1d | TRAPPIST | 0 | 7294.7736±0.0014 |
| | TRAPPIST | ? | 7367. 5818±0.0015 |

The transit timings shown were deduced from individual analyses of the transit light curves, assuming circular orbits for the planets. The error bars correspond to the 1$\sigma$ limits of the posterior PDFs of the transit timings.